\documentclass[a4paper,11pt]{article}
\pdfoutput=1 

\usepackage{jheppub} 

\makeatletter
\def\@fpheader{\relax}
\makeatother

\usepackage{graphicx,amsfonts,amssymb,amsmath}
\usepackage{bm}

\def\barray{\begin{eqnarray}}
\def\earray{\end{eqnarray}}
\def\beq{\begin{equation}}
\def\eeq{\end{equation}}

\title{A Gaussian Variational Approach to cMERA for Interacting Fields}

\author[a]{Jordan S. Cotler}
\author[b]{Javier Molina-Vilaplana}
\author[c]{Mark T. Mueller}

\affiliation[a]{Stanford Institute for Theoretical Physics, Stanford University, Stanford, CA 94305}
\affiliation[b]{Universidad Polit\'{e}cnica de Cartagena. C/Dr Fleming S/N. 30202 Cartagena. Spain}
\affiliation[c]{Center for Theoretical Physics, MIT, Cambridge, MA 02139 USA}

\abstract{We use the Gaussian variational principle to apply cMERA to interacting quantum field theories in arbitrary spacetime dimensions.  By establishing a correspondence between the first two terms in the variational expansion and the Gaussian Effective Potential, we can exactly solve for a variational approximation to the cMERA entangler.  As examples, we treat scalar $\varphi^4$ theory and the Gross-Neveu model and extract non-perturbative behavior.  We also comment on the connection between generalized squeezed coherent states and more generic entanglers.}

\begin{document} 
\maketitle
\flushbottom

\section{Introduction}
\label{sec:intro}
Tensor networks are a powerful tool for probing interacting many-body lattice systems.  They represent classes of systematic variational ansatzes which can be used in the Rayleigh-Ritz variational principle to solve for an approximation of an interacting ground state.  Methods also exist for determining excited states, finite temperature states, and time-dependent states.

A notable example in the tensor network literature is the multiscale entanglement renormalization ansatz, or MERA \cite{MERA1, MERA2, MERA3, Hauru1}.  This ansatz (or more accurately, class of ansatzes) provides a robust systematic variational approximation which has been demonstrated in many numerical studies \cite{nuMERAcle1, nuMERAcle2, nuMERAcle3, nuMERAcle4, nuMERAcle5, nuMERAcle6, nuMERAcle7, nuMERAcle8}.  The central insight of MERA is that it instantiates a modern version of Kadanoff's block spin renormalization group by creating states through building up entanglement correlations on a hierarchy of distance scales.  In classic block spin renormalization, one considers the effective Hamiltonian of blocks of spins on a lattice by coarse graining the Hamiltonian of the original spins.  This coarse graining procedure is iterated to capture the long-distance behavior of the system.  Similarly, the MERA ansatzes efficiently parametrize the effective entanglement between spins, blocks of spins, blocks of blocks of spins, and so on. 

It is natural to generalize MERA to continuum quantum field theories.  However, the continuum generalization known as cMERA has had comparatively limited scope, having only been applied to non-interacting field theories \cite{cMERA1, cMERA2, cMERA3}.
Furthermore, the standard cMERA ansatz lacks the generality of MERA, and has not been treated in any practical way as a systematic variational approximation like its lattice counterpart.

In this paper, we provide a first step towards applying cMERA to interacting QFT's in arbitrary spacetime dimensions.  We connect the methods of cMERA to existing work on variational methods in QFT, and suggest new connections with multi-mode generalized squeezed coherent states appearing in the quantum optics literature.

\section{From MERA to cMERA.}
\label{sec:MERA2cMERA}
To motivate the cMERA ansatzes, we first briefly review the MERA ansatz.  Suppose we are given a simple IR state $|\Psi_{\text{IR}}\rangle$ on a lattice, such as one with no entanglement that factorizes with respect to the lattice sites.  We would like to modify this state by building up entanglement at progressively smaller distances so as to create an effective state $|\Psi_u\rangle$ which captures correlations of the ground state of a Hamiltonian $H$ at a scale $u$ much less than the IR scale.

To create entanglement at each scale, we introduce the fine-graining isometries $W_i$ and the entangler unitaries $U_i$.  (In contexts where one instead flows from the UV to the IR, the Hermitian conjugates $W_i^\dagger$ and $U_i^\dagger$ are used, and referred to as coarse-grainers and disentanglers.)  A fine-graining isometry $W_i$ embeds a state into a larger lattice, effectively introducing new UV degrees of freedom.  In turn, an entangler unitary $U_i$ locally entangles the new UV degrees of freedom with the embedded degrees of freedom.  We construct the state $|\Psi_u\rangle$ by
\begin{equation}
|\Psi_u\rangle = U_u W_u \cdots W_1 U_1 W_0 U_0 |\Psi_{\text{IR}}\rangle \,,
\end{equation}
namely through applying fine-graining and entangling operations in succession at progressively smaller distance scales \cite{Hauru1}.  It is convenient to think of the combination $W_i U_i$ as implementing one step of the \textit{entangling process}, which requires both $U_i$ and $W_i$ applied in succession.  Then, we say that MERA performs the entangling process at successively smaller length scales.

MERA provides an efficient parametrization of the $W_i$'s and $U_i$'s, and allows one to economically minimize $\langle \Psi_u|\widehat{H}_u|\Psi_u\rangle$ with respect to those parameters, where $\widehat{H}_u$ is the Hamiltonian at scale $u$.  The state $|\Psi_u\rangle$, evaluated at the minimizing parameters, is then the variational approximation to the ground state of the system at scale $u$.  But since $|\Psi_u\rangle$ is built up hierarchically at different distance scales, the optimized ansatz also captures information about the entanglement of the ground state at all scales intermediate between $u$ and the IR.  Thus, while the variational principle minimizes the energy of the ansatz over short distance scales, the ansatz is hierarchically structured so that short-distance optimization leads to an accurate approximation of long-distance correlations.

The generalization to continuum QFT's is conceptually straightforward.  We write \cite{cMERA1, cMERA2, cMERA3}
\begin{equation}
\label{cMERAansatz1}
|\Psi_u\rangle = \mathcal{P} \, \exp\left[-i \int_{u_{\text{IR}}}^u (K(u') + L) \, du' \right] \, |\Psi_{\text{IR}}\rangle
\end{equation}
which contains the path-ordered exponential of the dilatation operator $L$ (which is the continuum analog of redefining the lattice spacing) and the generator $K(u')$ (which is the continuum analog of $W_{u'} U_{u'}$).  The path-ordering parameter $u'$ corresponds to the indices of $W_{u'} U_{u'}$, and is related to the scale at which we are performing the entangling process.  The entire path-ordered exponential is referred to as the entangler unitary.  Conventionally, the generator $K_{u'}$ is referred to as the cMERA entangler (or disentangler, depending on the direction of the RG flow we are performing), but this is unfortunate terminology: $K_{u'}$ should really be called the \textit{entangling processor} (or disentangling processor) since it corresponds to $W_{u'} U_{u'}$ and \textit{not} to $U_{u'}$ alone.  However, we will use the standard (albeit unfortunate) terminology of calling $K_{u'}$ the ``entangler'' in this paper.

Since the dilatation operator $L$ is fixed by the matter content of the QFT under consideration, the only variational parameters are those which parametrize $K(u')$.  At a conceptual level, we are done -- upon choosing the form of $K(u')$ we can attempt to solve the variational problem for the ground state of an interacting continuum QFT.  However, the art of variational methods lies in determining the particular form of the systematic variational approximation.  More specifically, while we have chosen a powerful hierarchical structure for our ansatz due to the form of Eqn. \!(\ref{cMERAansatz1}), it is impractical to optimize over \textit{all} Hermitian operator functions $K(u')$ and so we must choose a subspace (or sequence of subspaces) of Hermitian operator functions over which to optimize.

We will work with a QFT containing only one type of matter field, although the generalization to multiple matter fields is straightforward.  Let $[\phi(k), \pi(k')] = i\,\delta^d(k+k')$ and
\begin{align}
a_k &= \sqrt{\frac{\omega_k}{2}} \, \phi(k) + \frac{i}{\sqrt{2 \omega_k}} \, \pi(k) \\
a_k^\dagger &= \sqrt{\frac{\omega_k}{2}} \, \phi(k) - \frac{i}{\sqrt{2 \omega_k}} \, \pi(k)
\end{align}
where we use the notation $\omega_z := \sqrt{z^2 + m_0^2}$ with $m_0$ being the bare mass.  Notice that $[a_k, a^\dagger_{k'}] = \delta^d(k-k')$.
For applying the cMERA ansatz in Eqn.\! (\ref{cMERAansatz1}) to interacting QFT's in $d+1$ spacetime dimensions, one can expand $K(u')$ as
\begin{align}
\label{entangler1}
K(u') = &-\frac{i}{2} \int d^d k_1 \,\left(g_{1,0}(k_1;u') \, a_k^\dagger - \text{h.c.}\right) \nonumber \\
&- \frac{i}{2} \int d^d k_1 \int d^d k_2 \, \left(g_{2,0}(k_1, k_2; u') \, a_{k_1}^\dagger a_{k_2}^\dagger + g_{1,1}(k_1, k_2; u') \, a_{k_1}^\dagger a_{k_2} - \text{h.c.} \right) \nonumber \\
& - \frac{i}{2} \int d^d k_1 \int d^d k_2 \int d^d k_3 \, \left(g_{3,0}(k_1,k_2,k_3; u') \, a_{k_1}^\dagger a_{k_2}^\dagger a_{k_3}^\dagger + \cdots - \text{h.c.} \right) \nonumber \\
& - \,\, \cdots
\end{align}
where we will optimize over the variational functions $g_{i,j}$.  (Of course, one could instead expand $K(u')$ in $\phi(k)$ and $\pi(k)$ with related variational functions.) Note that in previous work as well as in this paper, $K(u')$ is approximated as at most quadratic in creation and annhilation operators \cite{cMERA1, cMERA2, cMERA3} (i.e., truncating Eqn. \!(\ref{entangler1}) after the first two lines) which can be viewed as the first few terms in a systematic expansion of $K(u')$.  Let us provide a physical interpretation of the expansion.  In Eqn. \!(\ref{entangler1}), we see that the quadratic terms are responsible for generating particle pair correlations as a function of scale, whereas the cubic terms generate particle triplet correlations as a function of scale, and so on.  Thus, truncating Eqn. \!(\ref{entangler1}) at order $n$ in the creation and annihilation operators means the truncated ansatz generates at most $n$-tuplet correlations at each scale.

It is clear that if we wish to describe the ground state of free field theory within the cMERA framework, then it is sufficient to truncate the expansion in Eqn. \!(\ref{entangler1}) at quadratic order, since a free theory is fully characterized by its connected 2-point functions.  Indeed, previous work on cMERA has only dealt with describing free fields using the quadratic form of $K(u')$.  In the remainder of this paper we develop techniques for \textit{analytically} solving the quadratic cMERA ansatz for interacting continuum QFT's in arbitrary dimensions.  We also speculate on techniques to move beyond quadratic order. \\

\section{Quadratic entangler.}
\subsection{Interacting scalar field theories.}
For the moment, let us work with relativistic, real scalar field theories with a mass gap.  Such theories flow to free fields in the IR, and thus their IR ground states are exactly Gaussian wavefunctionals.  We will minimize the expectation value of a Hamiltonian $\widehat{H}_u$ with respect to ansatz wavefunctionals $|\Psi_u\rangle$ which all have 1-point function $\langle \Psi_u|\widehat{\phi}(x)|\Psi_u\rangle = \overline{\phi}(x)$.  In the case $\overline{\phi}(x) = 0$, we are finding a variational approximation to the ground state of $\widehat{H}_u$ (if there is no spontaneous symmetry breaking).  Otherwise, for non-zero $\overline{\phi}(x)$, we are finding a variational approximation to the lowest-energy state of the theory with 1-point function $\overline{\phi}(x)$, which is an \textit{interacting} excited state.  Furthermore, optimizing over a class of states with fixed 1-point function gives us access to a variationally determined effective action $\Gamma[\overline{\phi}(x)]$ which generates correlation functions \cite{CJT1}.  In the remainder of this paper, we will consider the case that the $\overline{\phi}(x) = \overline{\phi}$ is constant, since our theories will be translation-invariant.

Letting $|\Omega\rangle$ be the scale-invariant IR state defined by \cite{cMERA2}
\begin{align}
\label{scaleinv1}
\left(\sqrt{C} \,(\phi(k) - \overline{\phi} \, \delta^d(k)) + \frac{i}{\sqrt{C}} \, \pi(k)\right) |\Omega\rangle = 0\, \quad \text{for all }k\,,
\end{align}
where $C$ is a constant to be determined later, and $\langle \Omega|\widehat{\phi}(x)|\Omega\rangle = \overline{\phi}$,
we can write down the most general cMERA ansatz with relativistic invariance that produces a translation-invariant 2-point function $G(k)$.  Truncating the entangler to quadratic order in creation and annihilation operators and working in the interaction picture with respect to the dilatation operator, we obtain the cMERA ansatz wavefunctional
\begin{align}
\label{cMERAwavefunctional1}
\Psi_u^{\text{scal.}}[\phi] &= \langle \phi| \, e^{-iuL} \, \mathcal{P} \, e^{\frac{1}{2}\int_{u_{\text{IR}}}^u du' \, \int d^d k \, \left(g(k e^{-u'}; u') \, a_k^\dagger a_{-k}^\dagger - g^*(k e^{-u'}; u') \, a_k a_{-k}\right) + \overline{\phi}\, \sqrt{\frac{\omega_0}{2}} \, (a_{0} - a_{0}^\dagger) }|\Omega\rangle \nonumber \\ \nonumber \\
&= \left[\det(2 \pi \,G)\right]^{-1/4} e^{-\frac{1}{4}\int^{\Lambda} d^d k \int^{\Lambda} d^d k' \, \left(\phi(k) - \overline{\phi}\right) \, G^{-1}(k ; u) \, \left(\phi(-k) - \overline{\phi}\right)} 
\end{align}
where the distance scale profile $g(k,u)$ factorizes as $g(k; u) = \chi(u) \cdot \Theta(|k|/\Lambda)$ and $\Theta$ is the Heaviside step function which implements a high-frequency cutoff at scale $\Lambda$.  Although we will work with a hard cutoff, the analysis generalizes simply for a soft cutoff function.  Let us unpack Eqn. \!(\ref{cMERAwavefunctional1}).  We have not included the number operator in the quadratic entangler ansatz since this will only contribute an overall phase.  The top line of the equation shows that the ansatz is a squeezed coherent state, with a multi-mode squeeze operator parametrized by $g(k; u')$ integrated over distance scales $u'$.  The bottom equation asserts that the cMERA ansatz is equivalent to a Gaussian wavefunctional with kernel $G^{-1}$ (i.e., the inverse Green's function) and mean field $\overline{\phi}$.  Supposing that $g(k,u)$ is purely real, one can use the QFT squeezed coherent state formalism \cite{Tsue1, Tsue2, Visser1} to show the equivalence
\begin{align}
\label{dictionary2}
G^{-1}(k;u_{\text{UV}}) &= 2\omega_\Lambda \, e^{-2 \int_{u_{\text{IR}}}^{u_{\text{UV}}} du' \, g(k e^{u'}; u')} \,,
\end{align}
where $\omega_\Lambda = \sqrt{\Lambda^2 + m_0^2}$ with $m_0$ being the bare mass.

In the case that our theory is a free massive scalar field, standard cMERA techniques allow one to solve for $g(k; u)$, and plugging this back into the ansatz gives the \textit{exact} ground state.  For interacting scalar field theories, solving the variational problem is more complicated.  Luckily, there is an extensive literature on performing variational optimization in an interacting QFT with a Gaussian wavefunctional ansatz \cite{GEP1, GEP2, GEP3, GEP4, Tsue1, Tsue2}.  This set of techniques goes by the name of the ``Gaussian Effective Potential."  Since the quadratic entangler cMERA ansatz is equivalent to the a Gaussian wavefunctional by Eqn. \!(\ref{cMERAwavefunctional1}), we can solve the variational problem in the Gaussian framework and then translate the answer into the cMERA framework via Eqn. \!(\ref{dictionary2}). 

Now let us give a concrete example of the variational principle in action for an interacting QFT.  For convenience, we take $u_{\text{IR}} = -\infty$ and $u = u_{\text{UV}} = 0$.  Consider real scalar $\varphi^4$ theory with a constant vacuum expectation value $\overline{\phi}$ and UV Hamiltonian
\begin{align}
\widehat{H}_\text{UV} = \int d^d x \, \left[\frac{1}{2} \widehat{\pi}^2 + \frac{1}{2}(\nabla \widehat{\phi})^2 + \frac{m_0^2}{2} \, \widehat{\phi}^2 + \frac{\lambda_0}{4!} \, \widehat{\phi}^4 \right] 
\end{align}
with an implicit high-frequency cutoff $\Lambda$.  Solving the variational problem, we find
\begin{align}
\label{gsol1}
g(k,u) =& \, \frac{1}{2} \, \frac{e^{2u}}{e^{2u} + M^2/\Lambda^2} \cdot \Theta(|k|/\Lambda) \\
\label{Csol1}
C =& \sqrt{\Lambda^2 + M^2}
\end{align}
where $C$ is a choice of the constant in the definition of $|\Omega\rangle$ in Eqn.~\eqref{scaleinv1}, and $M$ is the quasiparticle satisfying the gap equation
\begin{equation}
\label{gapeqn1}
M^2 = m_0^2 + \frac{\lambda_0}{2}\left(\overline{\phi}^2 + \int^{\Lambda} d^d k\, \frac{1}{2 \sqrt{k^2 + M^2}} \right)\,.
\end{equation}
One can show that the optimized ansatz captures all 1--loop $2$--point correlation functions, and additionally captures the resummation of all daisy and super-daisy diagrams \cite{GEP1}.  More generally, the variational optimization procedure is a machine which resums infinite classes of diagrams, allowing access to non-perturbative physics as per standard mean field theory \cite{GEP3}.

Notice that the only equation which remains to be solved to explicitly obtain the optimized ansatz is the gap equation (\ref{gapeqn1}).  Remarkably, wavefunctional optimization over an infinite-dimensional space of kernels $G$ has reduced to solving a single non-linear equation for $M$.  While it is easy to solve the gap equation (\ref{gapeqn1}) numerically, it can in many cases be solved analytically in the large $\Lambda$ limit.  For example, in 1+1 dimensions the solution is

\begin{equation}
\label{1plus1varmass}
M^2 = \frac{\lambda_0}{8\pi} \,\, W\left(\frac{32\pi\, \Lambda^2}{\lambda_0} \, e^{\frac{4\pi}{\lambda_0}(2 m_0^2 + \lambda_0 \overline{\phi}^2)}\right) \,\, + \mathcal{O}(1/\Lambda^2) \,,
\end{equation}
where $W(z)$ is the Lambert W function.  Indeed, Eqn. \!(\ref{1plus1varmass}) (and thus Eqn. \!(\ref{gsol1})) is non-perturbative in the coupling of the theory.

\subsection{Interacting fermionic field theories.}
For relativistic fermionic systems, Eqn. \!(\ref{cMERAwavefunctional1}) needs to be modified from the bosonic case although it will maintain a similar form.  One difference is that fermionic states have vanishing 1-point functions, and likewise so should the ansatz.  Letting $\theta(k), \theta^\dagger(k)$ be Grassmann-values functions following the formalism of \cite{JF1}, we have the fermionic wavefunctional ansatz

\begin{align}
\label{cMERAwavefunctional2}
\Psi_u^{\text{ferm.}}[\theta, \theta^\dagger] &= \langle \theta, \theta^\dagger | \, e^{-iuL} \, \mathcal{P} \,e^{\frac{1}{2} \int_{u_{\text{IR}}}^u du' \, \int d^d k \, \left(\psi_k^\dagger \,\textbf{g}(k e^{-u'}; u') \, \psi_{k} - \psi_k \, \textbf{g}^\dagger(k e^{-u'}; u') \,\psi_{k}^\dagger \right) }|\Omega\rangle \nonumber \\ \nonumber \\
&= \left[\det(4 \pi \,\textbf{G}^{-1})\right]^{1/4} e^{-\frac{1}{2}\int^{\Lambda} d^d k \int^{\Lambda} d^d k' \, \theta^\dagger(k) \, \textbf{G}(k, k'; u) \,\theta(k')} 
\end{align}
where in this case the distance scale profile $\textbf{g}(k; u)$ is a \textit{matrix} which factorizes as $\textbf{g}(k; u) = \boldsymbol{\chi}(u) \cdot \Theta(|k|/\Lambda) \, k$ and $|\Omega\rangle$ now denotes the fermion IR state.  Notice that $\textbf{g}(k; u)$ is now linear in $k$ so that the ansatz reproduces the exact ground state of a free fermion theory. Also, $\textbf{G}$ is now the fermionic Green's function.  We will restrict our attention to 1+1 dimensions where the gamma matrices are Pauli matrices, and the construction is particularly transparent.  We denote the fermion operator by $\psi = (\psi_1, \psi_2)$, with $\{\psi_1(k),\psi_1^\dagger(k')\} = \{\psi_2(k), \psi_2^\dagger(k')\} = \delta^d(k-k')$.  Furthermore, we use the IR state $|\Omega\rangle$ defined by \cite{cMERA2}
\begin{equation}
\psi_1(k) |\Omega\rangle = 0\,, \quad \psi_2^\dagger(k)|\Omega\rangle = 0\,, \quad \text{for all }k
\end{equation}
and ansatz
\begin{equation}
K(u) = i \int dk \left(g(k; u) \psi_1^\dagger(k) \psi_2(k) + g^*(k; u) \psi_1(k) \psi_2^\dagger(k)\right)
\end{equation}
where $g(k; u)$ is a scalar function.  We have the equivalence
\begin{equation}
\label{dictionary3}
\textbf{G}(k ;u_{\text{UV}}) = e^{2 \sigma_2 \int_{u_{\text{IR}}}^{u_{\text{UV}}} du' \, g(k e^{u'},u')}
\end{equation}
which is similar in form to Eqn. \!(\ref{dictionary2}).

As an example of the fermion formalism, consider the Gross-Neveu Hamiltonian in 1+1 dimensions, namely
\begin{equation}
\widehat{H} = \int dx \, \left[- i \widehat{\psi}^\dagger \alpha \cdot \nabla \widehat{\psi} - \frac{\xi^2}{2} (\widehat{\psi}^\dag \beta \widehat{\psi})^2 + \gamma \left(\sigma + \widehat{\psi}^\dagger \beta \widehat{\psi} \right) \right]
\end{equation}
with an implicit frequency cutoff $\Lambda$, where $\gamma \left(\sigma + \psi^\dagger \beta \psi \right)$ is a Lagrange multiplier term enforcing a bilinear condensate $\sigma$.  We can write $\alpha$ and $\beta$ in terms of Pauli matrices as $\alpha = (\alpha^1,\alpha^2) = (\sigma_1, \sigma_2)$ and $\beta = \sigma_3$.  Solving the variational problem, we find that the ground state comprises of (a linear combination of) two vacua with dynamically generated variational quasiparticle masses $M_{\pm}$ and non-zero condensate.  The entanglers for the two vacua have
\begin{equation}
\label{gsol2}
g_\pm(k,u) = \frac{1}{2} \bigg[-\arcsin \frac{\Lambda e^u}{\sqrt{\Lambda^2 e^{2u} + M_\pm^2}} + \frac{M_\pm \Lambda e^u}{M_\pm^2 + \Lambda^2 e^{2u}} \bigg] \cdot \Theta(|k|/\Lambda) \, k \,.
\end{equation}
The variational mass can be solved analytically for large $\Lambda$, and is given by
\begin{equation}
\label{GNvarmass1}
M_\pm = \pm \Lambda \cdot e^{-2\pi\left(\frac{1}{\xi^2} + I_0(\Lambda)\right)} \,\,+\mathcal{O}(1/\Lambda^2)\,.
\end{equation}
where the renormalized coupling $\xi_R$ at scale $\Lambda$ is
\begin{equation}
\xi_R(\Lambda)^2 = \left(\frac{1}{\xi^2} - I_0(\Lambda)\right)^{-1}\
\end{equation}
where we define
\begin{equation}
I_n(z) := \frac{1}{2}\, \int d^d p \, (p^2 + z^2)^{\frac{2n-1}{2}} \,.
\end{equation}
We see from Eqn. \!(\ref{GNvarmass1}) that $M_\pm$ is non-perturbative in the coupling $\xi$.

The techniques overviewed in this section generalize to systems of coupled bosons and fermions, gauge theories, and finite temperature theories \cite{OtherGEP1, OtherGEP2, OtherGEP3, OtherGEP4}.  Additionally, one can solve for time-dependent states including non-equilibrium states \cite{TimeDepGEP1, TimeDepGEP2, TimeDepGEP3, NonEquilGEP1}.  These same techniques are also useful for calculating variational approximations to entanglement entropy in interacting as well as quenched theories \cite{JM1, JM2, JM3}.

\subsection{Comments on the Renormalization Group.}

Above we have computed several examples of quadratic cMERA circuits.  To be precise, we have made two approximations: (1) the disentangler implementing the renormalization group (RG) flow is approximated as quadratic; and (2) the ground state wavefunctional is approximated as Gaussian.  These approximations are variational, and come from optimizing the cMERA ansatz with respect to the UV Hamiltonian.

Let us consider the $\varphi^4$ theory wavefunctional $\Psi_u^{\text{scal.}}[\phi]$ defined above for $d \leq 3+1$.  (see Eqn.'s~\eqref{scaleinv1},~\eqref{cMERAwavefunctional1},~\eqref{gsol1} and~\eqref{Csol1}).  As we tune the scale $u$ from $0$ (i.e., the UV scale) down to $-\infty$ (i.e., the IR scale), the quasiparticle mass $M$ obtained from $2$--point functions of $\Psi_u^{\text{scal.}}[\phi]$ becomes rescaled by $M \to e^{-u} M =: M(u)$.  Then the $\beta$--function of $M(u)$ is simply
\begin{equation}
\frac{dM(u)}{du} = - M(u) =: \beta(M(u))\,. 
\end{equation}
This is the same simple $\beta$--function as that of a free scalar field with mass $M$.  Of course, this is no surprise: we are \textit{approximating} the ground state of $\varphi^4$ theory as the ground state of a free massive scalar field theory with mass $M$, and so the RG flow of the mass is identical to that of a free massive scalar field theory.  Notice that the bare mass $m_0$ and bare coupling $\lambda_0$ are not themselves renormalized, but are merely part of the definition of the constant $M = M(0)$ due to the gap equation~\eqref{gapeqn1}.

Although the renormalization group of a free massive scalar field theory is simple, when cast in the cMERA formalism we are led to an interesting puzzle about the connection between cMERA and the Wilsonian renormalization group.  Suppose we have a free scalar field theory with UV Hamiltonian
\begin{equation}
\widehat{H}_\text{UV} = \int^\Lambda d^d k \, \left[\frac{1}{2}\, \widehat{\pi}(k) \, \widehat{\pi}(-k) + \frac{1}{2} \, \widehat{\phi}(k) (k^2 + m_0^2) \widehat{\phi}(-k) \right] \,,
\end{equation}
which is expressed in momentum space.  Its associated ground state is $|\Psi_{UV}\rangle$.  Indeed, the cMERA circuit $U$ can be obtained exactly in this case \cite{cMERA1, cMERA2}, and we can calculate the exact renormalized ground state (up to scale $\Lambda$) as
\begin{equation}
|\Psi_u\rangle = U^\dagger(0,u) |\Psi_{UV}\rangle\,.
\end{equation} 
for $-\infty < u \leq 0$.

A simple interpretation of Wilsonian RG would suggest that the renormalized Hamiltonian would be the effective Hamiltonian
\begin{equation}
\widehat{H}_{\text{eff}}(u) = \int^\Lambda d^d k \, \left[\frac{1}{2}\, \widehat{\pi}(k) \, \widehat{\pi}(-k) + \frac{1}{2} \, \widehat{\phi}(k) (k^2 + e^{-2u}\,m_0^2) \widehat{\phi}(-k) \right] 
\end{equation}
where the mass has been rescaled.  But in fact, cMERA gives that the renormalized Hamiltonian is $U^\dagger(0,u) \widehat{H}_\text{UV} U(0,u) = \widehat{H}_{\text{cMERA}}(u)$ which is only spatially \textit{quasi}-local \cite{cMERAVidal1}.  Notice that
\begin{equation}
\widehat{H}_{\text{eff}}(u) \not = \widehat{H}_{\text{cMERA}}(u)\,,
\end{equation}  
even though both Hamiltonians have the same, \textit{correct} renormalized ground state (up to scale $\Lambda$).  This seems to be in tension with na\"{i}ve expectations: one might expect the renormalized Hamiltonian to be strictly local, but contain renormalized parameters and irrelevant (albeit local) operators.  However, such a Hamiltonian could not be unitarily related to the UV Hamiltonian, since it would have a different spectrum.  In other words, to obtain a renormalized Hamiltonian that has the correct renormalized ground state \textit{and} the same spectrum as the UV Hamiltonian, the renormalized Hamiltonian may need to be only \textit{quasi}-local, as suggest by cMERA even in the free scalar field case.

\subsection{Beyond quadratic order.}
We briefly comment on techniques to go beyond quadratic order in the entangler, although explicit constructions will be given in future work.  The central difficulty in going beyond quadratic order is the ability to efficiently compute expectation values of a non-quadratic ansatz with respect to a Hamiltonian.  In particular, it is famously difficult to compute moments of non-Gaussian wavefunctionals.

Since the entangler $K(u')$ lies in a path-ordered exponential, one possible approach is to perform perturbation theory in its non-quadratic terms.  Although the resulting ansatz is perturbative in the idealized entangler, the resulting variationally optimized ansatz can be \textit{non}-perturbative in the parameters of the Hamiltonian, and indeed produce correlations which come from re-summing infinite classes of diagrams.  Various perturbative methods in this vein can be found in \cite{PostGaussian1, PostGaussian2, PostGaussian3, PostGaussian4, PostGaussian5, PostGaussian6}.

On the other hand, one could hope to gain analytic traction by choosing special forms of the disentangler which give rise to wavefunctionals with calculable moments.  Since the quadratic ansatz is a multi-mode squeezed coherent state, it is natural to understand its tractable generalizations which are discussed in the quantum optics literature (for example, see \cite{GeneralizedSC1, GeneralizedSC2, GeneralizedSC3, GeneralizedSC4}).  These techniques can yield a class of non-quadratic cMERA ansatzes with calculable moments.

\section{Discussion.}
\label{sec:discuss}
While cMERA has existed for a number of years, it has so far only been applied to free fields.  This provides a first step in applying cMERA to interacting fields, and widens the scope of cMERA's applications to include mean field theory.  We have shown that there is an equivalence between cMERA with a quadratic entangler and the Gaussian wavefunctional ansatz, and established a dictionary between the two variational methods for bosons and fermions.  Already at quadratic order in the disentangler, we showed that cMERA can access non-perturbative behavior of QFT's as per mean field theory.  In future work, it would be interesting to explore applications to holography and entanglement geometry, gapless theories like conformal field theories, and the landscape of generalized cMERA ansatzes.

\section*{Acknowledgements} 
JC is supported by the Fannie and John Hertz Foundation and the Stanford Graduate Fellowship program.  MTM is supported by the MIT Department of Physics under U.S. Department of Energy grant Contract Number DE-SC00012567. JMV is supported by Ministerio de Econom\'{i}a y Competitividad FIS2015-69512-R and Programa de Excelencia de la Fundaci\'{o}n S\'{e}neca 19882/GERM/15. We thank Patrick Hayden, Ali Mollabashi, Guifr\'{e} Vidal, Tom Imbo, Hong Liu, Nima Lashkari, Tobias Osborne and the referee for valuable conversations and feedback.

\end{document}